\documentclass[journal]{IEEEtran}

\usepackage{amsfonts}
\usepackage{amsmath}
\usepackage{booktabs}
\usepackage{multirow}
\usepackage{makecell}
\usepackage{graphicx}
\usepackage{textcomp}
\usepackage{array} 
\usepackage{longtable}
\usepackage{float}
\usepackage{caption}
\usepackage{subfigure}
\usepackage{doi}

\usepackage{enumitem}
\usepackage{hyperref}
\usepackage{cleveref}
\crefname{section}{Section}{Sections}
\usepackage{xcolor}

\graphicspath{{./figure/}}
\DeclareGraphicsExtensions{.pdf,.jpeg,.png,.jpg}

\ifCLASSOPTIONcompsoc
\usepackage[nocompress]{cite}
\else
\usepackage{cite}
\fi

\onecolumn
\title{Robust Multi-Read Reconstruction from Contaminated Clusters Using Deep Neural Network for DNA Storage}
\author{Yun Qin,
	Fei~Zhu~\IEEEmembership{Member,~IEEE}, and Bo Xi
    \thanks{This work was supported by the National Key Research and Development Program of China (No. 2020YFA0712100) .} 
	\thanks{The authors are with the Center for Applied Mathematics, Tianjin University, China. (~fei.zhu@tju.edu.cn) }
}

\begin{document}
	
\maketitle
\begin{abstract}
	DNA has immense potential as an emerging data storage medium. The principle of DNA storage is the conversion and flow of digital information between binary code stream, quaternary base, and actual DNA fragments. This process will inevitably introduce errors, posing challenges to accurate data recovery. 
	Sequence reconstruction consists of inferring the DNA reference from a cluster of erroneous copies. 
	A common assumption in existing methods is that all the strands within a cluster are noisy copies originating from the same reference, thereby contributing equally to the reconstruction. 
	However, this is not always valid considering the existence of contaminated sequences caused, for example, by DNA fragmentation and rearrangement during the DNA storage process.
	This paper proposed a robust multi-read reconstruction model using DNN, which is resilient to contaminated clusters with outlier sequences, as well as to noisy reads with IDS errors. 
	The effectiveness and robustness of the method are validated on three next-generation sequencing datasets, where a series of comparative experiments are performed by simulating varying contamination levels that occurring during the process of DNA storage. 
	
\end{abstract}

	
\section{Introduction} ~\label{sec1}
\begin{figure*}[htbp]
	\centering
	\includegraphics[scale=0.2]{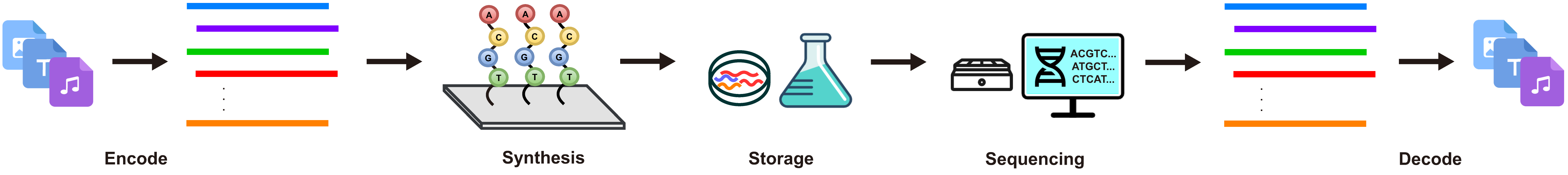}
	\captionsetup{labelfont=bf}
	\caption{Overview of the DNA storage system. The workflow consists of five stages: encoding, synthesis, storage, sequencing, and decoding.}
	\label{fig:1}
\end{figure*}

\IEEEPARstart{N}{owadays}, the information explosion leads to the generation of massive data, that brings great challenges to traditional storage systems, such as mobile hard disks, USB flash memory, and integrated circuits. When utilizing these storage mediums, several problems arise inevitably, including insufficient storage duration, high energy consumption, and environmental pollution~\cite{goda2012history}.
Meanwhile, Deoxyribonucleic Acid (DNA) molecule emerges as a promising storage medium, owing to its theoretically high storage density and long storage term, which fits the request of storing huge amounts of data~\cite{zhirnov2016nucleic, ceze2019molecular}. The workflow of DNA storage is summarized in Figure \ref{fig:1}. 


Generally, the DNA storage consists of firstly encoding binary stream to the alphabet \{A, T, C, G\} strings, chemically synthesizing short DNA oligos, namely {\em references}, and then storing the synthesized DNA strands  {\em in vitro} or {\em in vivo}.
To read the information via next-generation sequencing, the references should be retrieved from a large, unordered collection of error-prone {\em reads}. 
This is because both synthesis and sequencing in DNA storage inevitably introduce insertion-deletion-substitution (IDS) errors to the DNA strands, with the error probability being 1\%-2\% in the mainstream next-generation sequencing and up to 10\% for Nanopore sequencers \cite{dong2020dna}. 
During sequencing, each single reference outputs an uncertain number of noisy copies, and the reads corresponding to different references are gathered without ordering~\cite{zhirnov2016nucleic, meiser2020reading}. 
 {\em Clustering} is usually applied on the sequencing file, such that the noisy reads originated from the same reference are grouped into clusters~\cite{rashtchian2017clustering}. After that, the {\em multi-read reconstruction}, which is the topic of this paper, is performed to infer the the original reference from a cluster of noisy reads~\cite{sabary2020reconstruction}.


During the past ten years, a lot of research has been devoted to the sequence reconstruction problem in DNA storage. Roughly, they are divided into three categories: the consensus methods of  Bitwise Majority Alignment (BMA)~\cite{gopalan2018trace, yekhanin2020trace, srinivasavaradhan2019symbolwise, sabary2020reconstruction}, the statistical inference methods~\cite{shibata2016fixed,  sakogawa2020symbolwise, lenz2021concatenated}, and the recent deep learning ones~\cite{bar2021deep, nahum2021single, lv2020end}. 
The BMA and its variations are elaborated for IDS channels and applied to DNA storage systems in~\cite{gopalan2018trace, yekhanin2020trace, sabary2020reconstruction}. 
They perform position-to-position alignment among multiple reads and implement a majority voting strategy. The BMA-based methods are effective, especially for datasets with low IDS error rates. 

The second category is based on statistical inference, where at each position of the sequence, the maximum a posterior (MAP) probabilities of all the possible input symbols are estimated and compared~\cite{shibata2016fixed, sakogawa2020symbolwise, lenz2021concatenated}. 
In~\cite{shibata2016fixed}, marker codes are inserted into LDPC codes at fixed intervals for error correction, and the decoder is based on a forward and backward (FB) algorithm. 
In~\cite{sakogawa2020symbolwise}, a drift vector is introduced to model the insertion/deletion errors in each received word, and a factor graph is derived for joint probability estimation. 
Concatenated codes are considered in~\cite{lenz2021concatenated}, whose inner codes and channels are modeled as joint Hidden Markov Models (HMM) and the BCJR inference is derived. 
The so-called Trellis BMA marries BMA with BCJR decoding and achieves a linear complexity in the number of traces~\cite{srinivasavaradhan2021trellis}. 
However, due to the computational overhead, the feasible reads number per cluster can hardly exceed ten when applying these methods in practical DNA storage systems.  

With the emergence of deep learning, a few lately works have attempted to exploit deep neural networks (DNN)  to address the multi-read reconstruction~\cite{bar2021deep}, as well as single read reconstruction~\cite{nahum2021single} in DNA storage systems.
Similar in spirit of this work, the main idea is to train a DNN model with good error correction capacity, that can map a cluster of noisy reads to the corresponding DNA reference. As this work also focuses on the multi-read reconstruction using DNN,  the relevant works~\cite{bar2021deep, nahum2021single, lv2020end} will be reviewed in~\cref{sec2}.

\begin{figure}[htbp]
	\centering
	\includegraphics[scale=0.4]{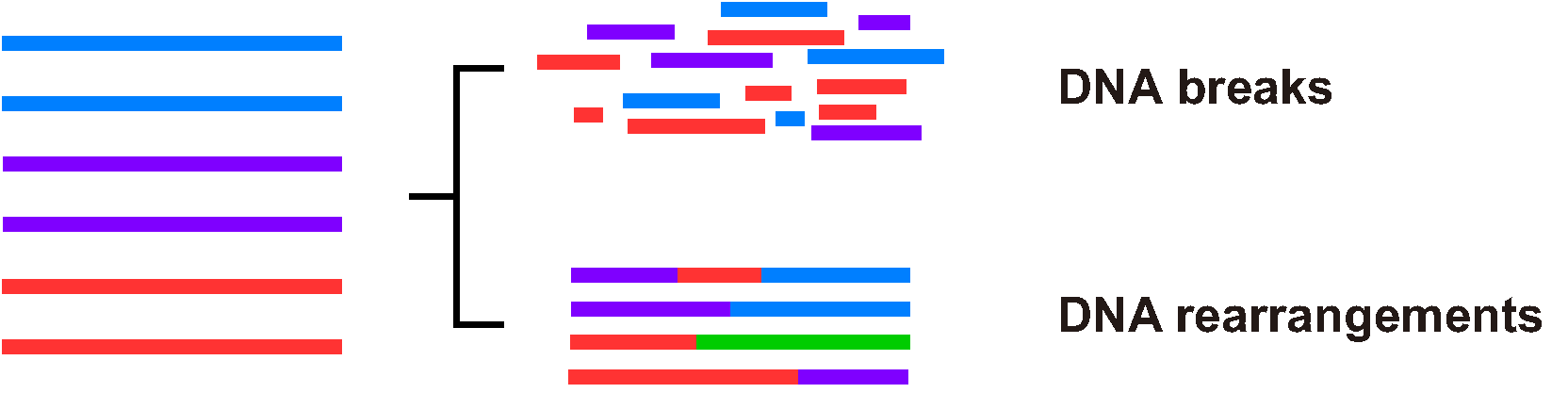}
	\captionsetup{labelfont=bf}
	\caption{Illustration of strands breaks and rearrangements in DNA data storage.}
	\label{fig: contaminated}
\end{figure}

In practice, the stability and robustness of current DNA storage systems are threatened by contaminated sequences that occur at different stages of the DNA storage. 
Unlike a noisy read that differs from its reference by only a few IDS errors, we refer the contaminated sequences to strands with a more significant edit distance from the original DNA references. 
Several factors contribute to the occurrence of contaminated sequences. 
In long-term storage and under certain conditions, DNA strands are susceptible to degradation, which results in strand breaks and loss~\cite{matange2021dna}. 
Unspecific amplification inevitably causes frequent DNA breaks and rearrangements, where oligos are fragmented and rejoined to new ones~\cite{Song2022}, as shown in Figure \ref{fig: contaminated}.  
Contaminated sequences also include the complementary strands of the references produced during sequencing~\cite{mallet2021reverse}. 
Considering the security issue in DNA storage, contaminated sequences are intentionally added for the purpose of data encryption in~\cite{kim2020metastable, shomorony2021dna, vippathalla2022secure}. 
Obviously, the existence of contaminated sequences makes the already challenging reconstruction problem more difficult~\cite{matange2021dna, Song2022}. 

In all aforementioned methods, every strand within a cluster contributes equally to the reconstruction of the reference strand, which holds only when the cluster under reconstruction comprises only the noisy copies originating from the same reference. 
However, such prerequisite for perfect clustering is not always achievable, accounting for the properties of current DNA storage systems. 
When the sequencing file contains a portion of contaminated sequences, clustering algorithms will fail to generate clusters in accordance with latent DNA references. 
On the other hand, as sequencing are biased towards strands with specific properties, existing perfect clustering methods ({\em e.g.},~\cite{zorita2015starcode, rashtchian2017clustering, qu2022clover}) have the risk of losing references rarely sequenced~\cite{bar2021deep, matange2021dna}.  
That is, clustering them into the wrong clusters.  
To the best of our knowledge, there is no method to differentiate the sequence quality and reliability within the cluster, in the context of sequence reconstruction for DNA storage.

This paper proposes a robust multi-read reconstruction method based on DNN. Taking advantage of the attention mechanism and the conformer block, the proposed model is resilient to contaminated clusters with outlier sequences, as well as noisy reads with IDS errors. The main contributions are as follows:
\begin{itemize}
	\item{}\textbf {Integration of sequence quality to multi-read reconstruction.} By far, this is the first multi-read reconstruction method that takes into account sequence reliability within the cluster. After scored according to sequence quality by the attention module, strands will contribute to the reconstruction at varying degrees. Thus the effect of various kinds of contaminated sequences can be suppressed automatically. 
	\item{}\textbf {Error correction capacity of IDS errors within cluster.} The proposed model realizes the error correction of IDS errors within the cluster. The Conformer-Encoder has strong feature extraction ability, such that the local features extracted by the convolutional layers and global features extracted from the attention module are smartly integrated. The resulting features are high-level and representative, such that the underlying reference of the noisy cluster can be well recovered by a single-layer long short-term memory (LSTM) decoder.
	\item{}\textbf {Sequence reconstruction model accommodating varying cluster sizes.} The network is trained directly from clusters of different sizes, rather than summing up the reads within a cluster to form a structured input format~\cite{bar2021deep}. Thereby, it is compatible with the input cluster of varying sizes at the testing stage.  
	\item{}\textbf {Small network with less parameters.} The proposed neural network has a small structure ($\approx 2.5 $ M parameters) with good generalization ability. This helps to mitigate the overfitting issue caused by the shortage of training data, when using DNN to address the sequence  reconstruction problem in DNA storage. 
\end{itemize}

The rest of this paper is organized as follows. The related work is reviewed in~\cref{sec2}. In~\cref{sec3}, we present the proposed multi-read reconstruction model. Experimental results and analysis are given in~\cref{sec4}. Finally, ~\cref{sec2} concludes the paper.

\section{Related Work} ~\label{sec2}
We succinctly review several deep learning-based sequence reconstruction methods in DNA storage. 
The most relevant literature to this paper is the so-called DNAformer, a scalable and robust solution for the DNA sequence reconstruction recently proposed in~\cite{bar2021deep}. The model is based on DNN, and is well adapted to imperfect but fast clustering of copies. 
Benefiting from convolution, Xception and transformer, the model has good capacity to correct IDS errors (especially substitutions) within the cluster. 
Besides dissimilar network designs, our method differs from DNAformer in the following aspects.
\begin{enumerate}[leftmargin=*]
\item The input of DNAformer is an element-wise sum of multiple copies, implying every sequence within a cluster equally important to the reconstruction.
It fails to consider the differences in sequence quality caused by the existence of contaminated sequences. 
On the contrary, our method scores every sequence within the cluster, and accordingly, the strands contribute to the reconstruction at different levels. 
\item To overcome the shortage of training data, DNAformer applies the Synthetic Data Generator (SDG)~\cite{SDG} to generate sufficient DNA sequences for training the model, with the sequence error rates estimated by SOLQC~\cite{btaa740}. 
Alternatively, our method circumvents this issue by designing a small but efficient network, which can be trained with much fewer labeled samples.   
\end{enumerate}

Nahum \textit{et al.} \cite{nahum2021single} established a single-read reconstruction model for DNA based storage systems, aiming at understanding the error patterns from only a single sequence by a global, context-aware method. 
The model uses an encoder-decoder transformer architecture composed of two paired BERT models. 
The error correction is regarded as a self-supervised sequence-to-sequence task, and the network is trained using synthetic sequences generated by SDG \cite{SDG}.


In~\cite{lv2020end}, a basecaller-decoder integration method is proposed for recovering Nanopore sequencing data, where the Viterbi error correction and recurrent neural network (RNN) are combined.  
The reference is reconstructed directly from multiple records of the raw signal, instead of being inferred from highly noisy basecalled reads. This yields a 3-fold reduction in reading cost compared to previous work.
%
%

\begin{figure*}[htbp]
	\centering
	\vspace{-0.1cm}
	\includegraphics[width=\textwidth]{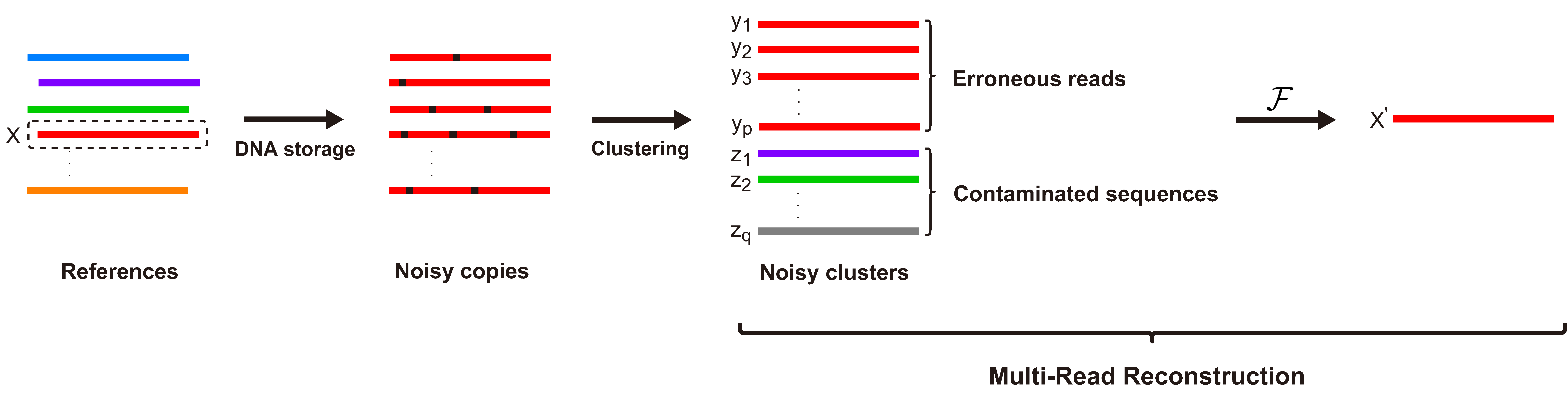}
	\captionsetup{labelfont=bf}
	\caption{Illustration of the multi-read reconstruction problem defined in this paper. Binary files are encoded as DNA references. 
	Multi-read reconstruction starts from a noisy cluster containing the erroneous copies originating from the original reference and the contaminated sequences occurring at different stages of DNA storage. The proposed reconstruction method aims at finding a mapping (characterized by a neural network), that minimizes the distance between the cluster and the original reference.}
	\label{fig:2}
\end{figure*}

\section{The Proposed Model}~\label{sec3}
We formulate the multi-read reconstruction problem mathematically, and then describe the proposed  Robust multi-read Reconstruction from
Contaminated Clusters using DNN (RRCC-DNN) method in detail.

\subsection{Problem Statement}~\label{problem}
Use ${\Sigma = \lbrace A, C, G, T\rbrace}$ to represent four DNA nucleotides.
Let $\mathcal{C}\in ({\Sigma}^{\ast})^{N}$ be a noisy cluster, which contains $p$ erroneous reads $y_1, y_2, ... , y_p$ originating from the same reference $x \in{\Sigma}^{L}$, and $q$ contaminated sequences $z_1, z_2, ... , z_q$ introduced at various stages of the DNA storage process, with $N=p+q$. 
Based on this assumption, the DNA multi-read reconstruction algorithm is a mapping: 
\begin{equation}
	\nonumber
	\mathcal F: \mathcal{C} \rightarrow \Sigma^{L},
\end{equation}
which receives $N$ sequences and outputs $\hat x$, an estimate of $x$, as shown in Figure \ref{fig:2}.
In this work, we focus on deploying a DNN model to find such a mapping $\mathcal F$ that the distance between $\hat x$ and $x$ can be minimized. 

\begin{figure*}[htbp]
	\centering
	\vspace{-0.2cm}
	\includegraphics[width=.85 \textwidth]{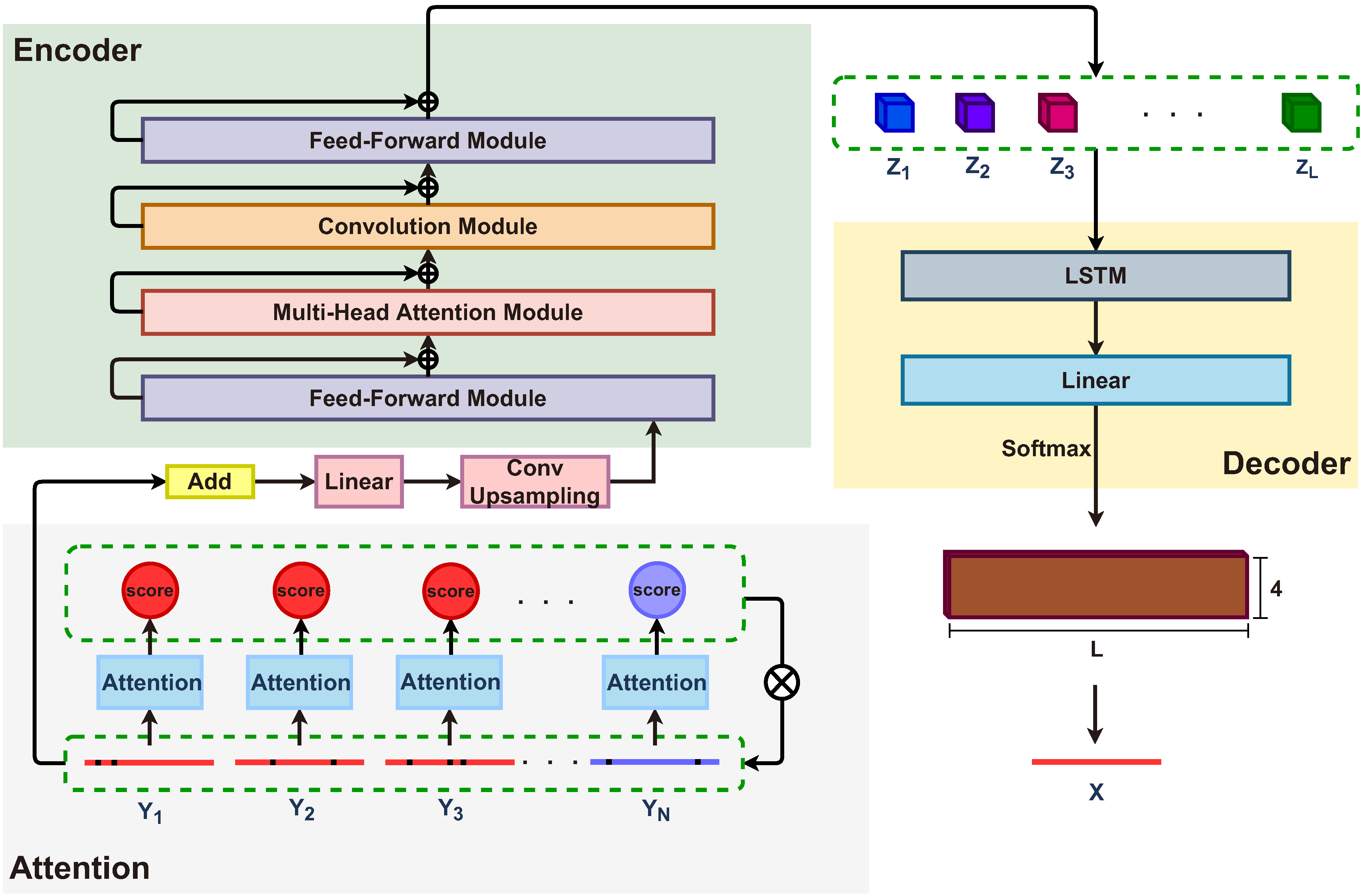}
	\captionsetup{labelfont=bf}
	\caption{Model architecture. The proposed RRCC-DNN is composed of Attention Module, Conformer-Encoder, and LSTM-Decoder, which correspond to the three colored regions in the figure, respectively.}
	\label{fig:model}
\end{figure*}

\subsection{Model Overview}
We aim at addressing the sequence reconstruction problem defined in~\cref{problem} by deep learning.
As shown in Figure~\ref{fig:model}, the proposed neural network is based on the encoder-decoder architecture, and is mainly composed of three components, {\em i.e.}, Attention Module, Conformer-Encoder and LSTM-Decoder. 
The attention mechanism~\cite{chandak2019improved} is used to automatically suppress the effect of suspicious contaminated sequences while amplifying the contribution of sequences that likely originating from the cluster reference.  
Placed at the front of the model, the Attention Module scores the quality of every sequence of the input cluster, and generates a high-level, average-weighted feature accordingly. 
The Conformer-Encoder is expected to understand the IDS error patterns within a cluster, taking into account its powerful feature extraction ability. 
It interactively combines the local features extracted by the convolution with the global features generated by the attention module. 
The decoder is a single-layer LSTM, which outputs the predicted reference of the input cluster. 
Next, we present the sequence embedding and three model components in detail, as well as the loss function. 
\subsubsection{Sequence Embedding}
The model input is a cluster of a non-fixed number of DNA sequences with varying lengths. 
Before being fed to the network, each sequence is represented by the one-hot encoding to a prefixed, uniform length $L$, where zeros are padded to short strands. 
In this way, every sequence is converted to a matrix of size $4\times L$, each column being a one-hot vector indicating the corresponding base at that index position. 


\subsubsection{Attention Module}
\begin{figure*}[htpb]
	\centering
	\includegraphics[width=.85\textwidth]{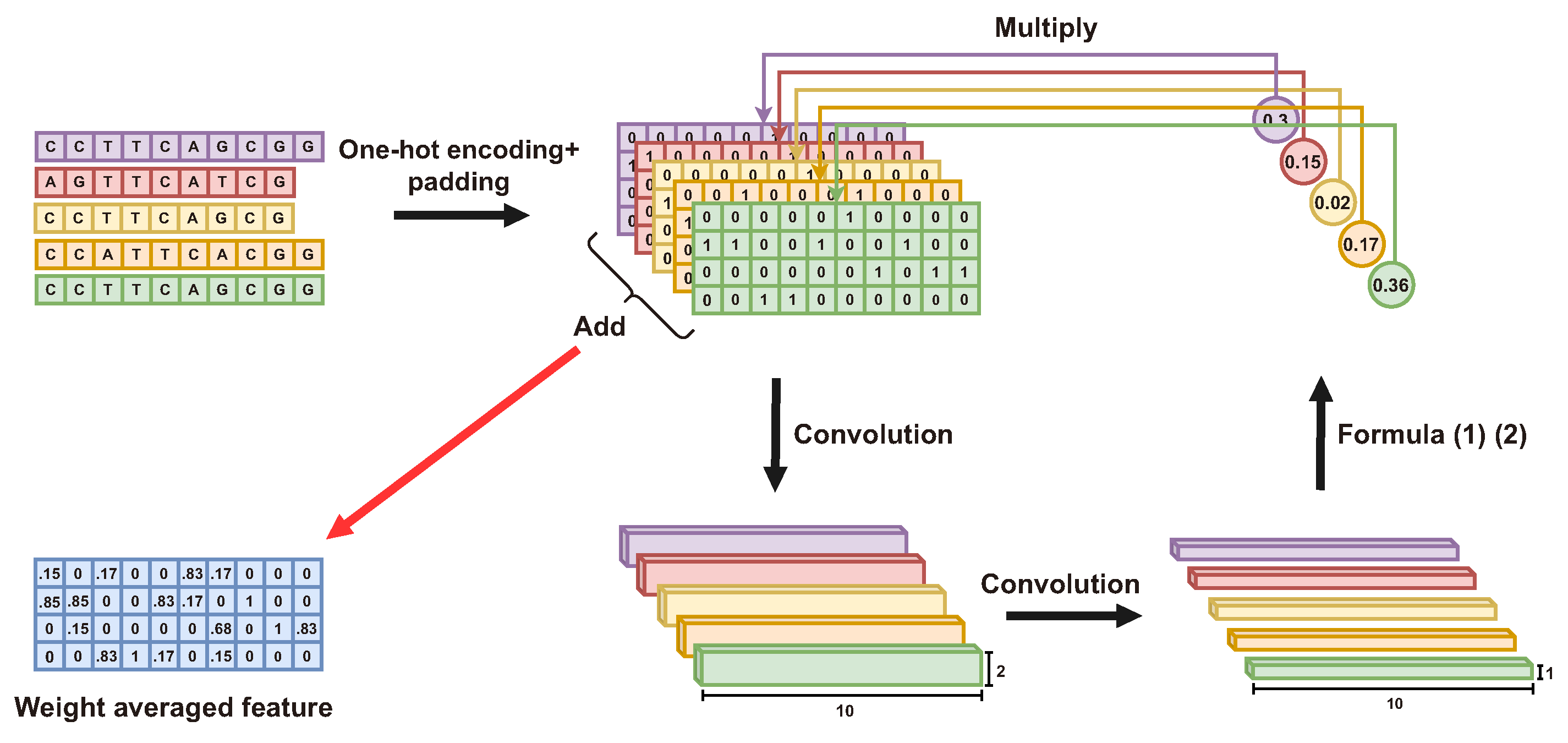}
	\captionsetup{labelfont=bf}
	\caption{Attention Module. Each noisy copy is converted to a two-dimensional matrix by one-hot encoding and zero-padding. After convolution, each feature is transformed into a one-dimensional vector, and is fed to the attention mechanism to estimate a scalar score. Finally, the weight-averaged feature for the given cluster is generated.}
	\label{fig:attention}
\end{figure*}

As illustrated in Figure \ref{fig:attention}, the attention module consists of the convolutional layer followed by an attention mechanism~\cite{vaswani2017attention}. 
For every strand feature, we perform two successive 1D convolution operations with kernel sizes of 3 and 5 to model the position shifts from synchronization errors, while reducing the number of feature channels from 4 to 2 and finally to 1. 
The resulting one-dimensional vectors are scored by the attention mechanism in a similar way as in~\cite{desplanques20_interspeech}.
%

Let $y_i \in \mathcal{R}^{L \times 4}$ be the input feature of the $i$-th strand in cluster, and  
$\widetilde{y}_i \in \mathcal{R}^L$ be the corresponding vector after convolution. 
The attention mechanism is applied as
\begin{equation}
	e_i=v^Tf(W \widetilde{y}_i+b)+k.
\end{equation}
Here, the linear transform with parameters $W$ and $b$ is used to project the vector to lower-dimensional space, thus reducing parameter number of the network.
After a nonlinear activation layer $f$, the feature is transformed to a sequence-wise attention score $e_i$ via a linear layer (parameterized by $v$ and $k$). 
By applying the softmax function, the scalar $e_i$ is normalized over all the strands within the cluster as
\begin{equation}
	\alpha_i=\frac{\exp(e_i)}{\sum_{i=1}^N{\exp(e_i)}},
\end{equation}
where $N$ is the cluster size, and $\alpha_i$ is the final attention score of the $i$-th sequence.
Obviously, the attention score reflects the importance of each strand within the cluster. 
As a result, the weight-averaged feature for the given cluster becomes 
\begin{equation}\label{feature}
	y =\sum_{i=1}^N{{\alpha}_i y_i},
\end{equation}
Here, every sequence contributes to the representation differently according to sequence quality, with the importance of high-quality reads amplified and the effect of low-scored strands suppressed automatically. 
After the attention module, the linear layer and convolution upsampling is applied to represent the feature~\eqref{feature} in an enlarged feature space. 
As a 4-dimensional representation is not enough to characterize a position in the sequence, we expand the feature dimension from $L \times 4$ to  $L \times 128$ in the experiments. 

\subsubsection{Conformer-Encoder}
Concerning the encoder, we adapt the convolution-augmented transformer (conformer), which is proposed for speech recognition and outperformed the CNN and transformer-based models with state-of-the-art results~\cite{gulati2020conformer}. The conformer combines self-attention and convolution, where the former captures the global feature while the latter learns the relative-offset-based local interactions. As a result, it can correct IDS errors through the extracted rich semantic information.
As shown in Figure~\ref{fig:model}, Conformer-Encoder consists of multi-head self-attention layers and convolution layers sandwiched between two feed-forward modules with shortcut connections, where  layer normalization is always applied at the junction of two modules. 

The multi-head attention module (\rm{MHSA})~\cite{vaswani2017attention} is computed by scaled-dot product with 
\begin{equation}\label{self-attention}
	\rm{Attention}(Q,K,V)=\rm{Softmax}(\frac{Q K^\top}{\sqrt{d_k}})V,
\end{equation}
where $Q, K, V \in{\mathcal{R}^{L\times d_k}}$ are linear transformations of feature $X\in \mathcal{R}^{L\times d}$.
In this work, we employ $h=8$ parallel attention heads, namely the concatenation of $h$ scaled-dot product attention results, yielding
\begin{equation}
	\rm{MHSA}(X)=\rm{Concat}(\rm{head_1,head_2,...,head_h})W,
\end{equation}
where $head_i$ is computed from~\eqref{self-attention}, and $W\in{\mathcal{R}^{hd_k\times d}}$ maps the concatenated feature back to the original dimension $d$.
In practice, we have $d_k=128$ and $d=d_k/h=16$.
As for the convolution module (Conv), we perform two deep separable convolutions with kernel size 31 to capture local correlations among sequence positions. 
Each feed-forward module (FFN) has two linear layers, which firstly double and then restore the original feature dimension. 

Mathematically, for $x$ input to Conformer-Encoder, the output $y$ is:

\begin{equation}
	x^{'}=x+\frac{1}{2} \rm{FFN}(x)
\end{equation}

\begin{equation}
	x^{''}=x^{'}+\rm{MHSA}(x^{'})
\end{equation}

\begin{equation}
	x^{'''}=x^{''}+\rm{Conv}(x^{''})
\end{equation}

\begin{equation}
	y=x^{'''}+\frac{1}{2} \rm{FFN}(x^{'''})
\end{equation}

\subsubsection{LSTM-Decoder}
As an advanced variant of RNN, LSTM can model long-range decencies well for chronological data \cite{7508408}.
As DNA is context-dependent and sequential data, we employ a single-layer LSTM decoder. 
Although simple, it is sufficient for the reconstruction task, owing to the powerful feature extraction ability of Conformer-Encoder.
The decoder reduces the feature dimension back to 4, outputting for each position the estimated probabilities for each base.

The proposed RRCC-DNN model is trained using a cross-entropy loss function defined as
\begin{equation}
	\mathcal{L}=-\sum_{l}^{L} x_{l}\log f(y_{l}), 
\end{equation}
where $L$ is the sequence length, $y_{l}$ is one-hot label vector indicating the base category for the $l$-th position, and $f(x_{l})$ represents the predicted probability vector by the proposed neural network.

\begin{table*}[htbp]
	\centering
	\captionsetup{labelfont=bf}
	\caption{Data description.}
	\label{tab: data describe}
	\begin{tabular}{m{25em}ccc}
		\toprule
		\textbf{Dataset} \centering & \multicolumn{1}{c}{\textbf{Erlich \textit{et al.}\cite{erlich2017dna}}} & \multicolumn{1}{c}{\textbf{Organick \textit{et al.}\cite{organick2018random}}} & \multicolumn{1}{c}{\textbf{Chandak \textit{et al.}\cite{chandak2019improved}}} \\
		\midrule
		\textbf{Number of original sequences} \centering & \multicolumn{1}{c}{ 72000} & \multicolumn{1}{c}{ 607150} & \multicolumn{1}{c}{ 11710} \\
		\textbf{Length of the original sequence} \centering & \multicolumn{1}{c}{ 152} & \multicolumn{1}{c}{ 150} & \multicolumn{1}{c}{ 150} \\
		\textbf{Synthesis} \centering & \multicolumn{1}{c}{ Twist Bioscience} & \multicolumn{1}{c}{Twist Bioscience } & \multicolumn{1}{c}{CustomArray} \\
		\textbf{Sequencing} \centering & \multicolumn{1}{c}{ Ilumina miSeq} & \multicolumn{1}{c}{Ilumina NextSeq } \centering & \multicolumn{1}{c}{Illumina iSeq } \\
		\textbf{Number of original sequences aligned to reads} \centering & 72000 & 596669 & 11710 \\
		\textbf{Missing clusters} \centering & 0     & 10481 & 0 \\
		\textbf{Number of reads aligned to original sequences} \centering & 13328870 & 14486345 & 1065117 \\
		\bottomrule
	\end{tabular}
\end{table*}

\begin{table*}[htbp]
 \centering
 \captionsetup{labelfont=bf}
 \caption{Statistics of the training and testing set.}
 \label{tab:data statistic}
 \renewcommand{\arraystretch}{1.3}
 \begin{tabular}{c c c c c}
  \toprule
&{\textbf{Dataset}} & \multicolumn{1}{c}{\textbf{Erlich \textit{et al.} \cite{erlich2017dna}}} & \multicolumn{1}{c}{\textbf{Organick \textit{et al.} \cite{organick2018random}}} & \multicolumn{1}{c}{\textbf{Chandak \textit{et al.} \cite{chandak2019improved}}} \\
  \midrule
\multirow{3}{*}{\textbf{Training Set}} 
&{\textbf{ cluster number } }& \multicolumn{1}{c}{ 36000} & \multicolumn{1}{c}{ 296317} & 5857 \\
&{\textbf{cluster size}} & \multicolumn{1}{c}{ 5-30} & \multicolumn{1}{c}{ 5-30} & \multicolumn{1}{c}{ 5-30} \\
&{\textbf{number of reads}} & \multicolumn{1}{c}{ 628875} & \multicolumn{1}{c}{ 5587728} & 101643 \\
\hline
\multirow{3}{*}{\textbf{Testing Set}} 
&{\textbf{ cluster number}} & \multicolumn{1}{c}{ 36000} & \multicolumn{1}{c}{ 296325} & 5853 \\
&{\textbf{cluster size}} & \multicolumn{1}{c}{ 5-30} & \multicolumn{1}{c}{ 5-30} & \multicolumn{1}{c}{ 5-30} \\
&{\textbf{ number of reads}} & \multicolumn{1}{c}{ 630945} & \multicolumn{1}{c}{ 5586351} & 102744 \\
  \bottomrule
 \end{tabular}
\end{table*}
\section{Experimental Results}~\label{sec4}
\subsection{Data Preparation and Training details}
We use three well-known datasets for DNA-based storage provided in Erlich et al.\cite{erlich2017dna}, Organick et al.\cite{organick2018random}, and Chandak et al.\cite{chandak2019improved}. 
Dataset descriptions are given in Table \ref{tab: data describe}. 
Each dataset comprises two files for sequence reconstruction, one containing the disordered collection of the noisy reads and the other recording all the original references. 

As no ground-truth clusters are available in practice, we first apply Burrows-Wheeler-Alignment Tool (BWA)~\cite{li2009fast} on both files, and 
take the sequence alignment results as perfect clusters, where each read is matched to its closest reference.   

In the experiments, we set the cluster scale to be 5$\sim$30, a modest range for the sequence reconstruction task, by randomly picking reads from each previously-obtained cluster.
It is because the number of reads in the original sequencing file is large, signifying a commonly large cluster size with much information redundancy ({\em e.g.}, averaged number of copies per reference $\approx 185$ in dataset {{Erlich \textit{et al.}\cite{erlich2017dna}}, Table \ref{tab: data describe}). 

To simulate the challenging scenarios where DNA storage is under threat of contamination, we inject into each cluster a certain proportion of contaminated sequences generated from the following reasons:
\begin{itemize}
	\item{} {\textbf{Misclustered sequence.} When the clustering is imperfect, a sequence will be assigned to the wrong cluster. }
	\item{} {\textbf{Reverse complementary strand.} The sequencing process generates the reverse complementary sequence for a DNA~\cite{mallet2021reverse}, and the strands in opposite orders risk coming to the same cluster with imperfect clustering.}
	\item{} {\textbf{Random sequence.} Inspired by \cite{kim2020metastable}, we consider the randomly generated DNA to simulate the fake information intentionally added to the original file.}
\item{} {\textbf{Splicing of DNA fragments.} Such errors are simulations of DNA breakages and rearrangements that frequently occur in DNA storage and PCR amplification-based DNA strand replication, as mentioned in \cite{Song2022}. }
\end{itemize}
To demonstrate the effectiveness of the proposed model under different contamination levels,  we inject contaminated sequences into each cluster, with equal probability for every candidate reason. 
More precisely, on each of the three datasets, five simulations with contamination levels ranging within the set $\{0\%, 5\%, 10\%, 15\%, 20\%\}$ are performed by using the proposed method as well as the comparative sequence reconstruction approaches.
Here, $0\%$ corresponds to the case without extra added contaminated sequences, and the clusters are composed of the reads from the original sequencing file. 

Table~\ref{tab:data statistic} reports the training and testing set on three datasets.
For each experiment, the proportion of training data to test data is set to 1:1. 
We sort the training samples according to their cluster sizes and divide the training set into batches such that each comprises clusters of the same size. 
The training and testing are performed on a single 2080ti GPU. 
We set the batch size to 64 and the initial learning rate to 0.005. 
The Adam optimizer is applied with parameter values $\beta_1=0.9$ and $\beta_2=0.98$. The coefficient of $L_2$ regularization is chosen as $1e-4$ to prevent model overfitting. 

\begin{figure}[htbp]
	\centering
	\subfigure[Erlich \textit{et al.}\cite{erlich2017dna}]{
		\includegraphics[width=0.5\textwidth]{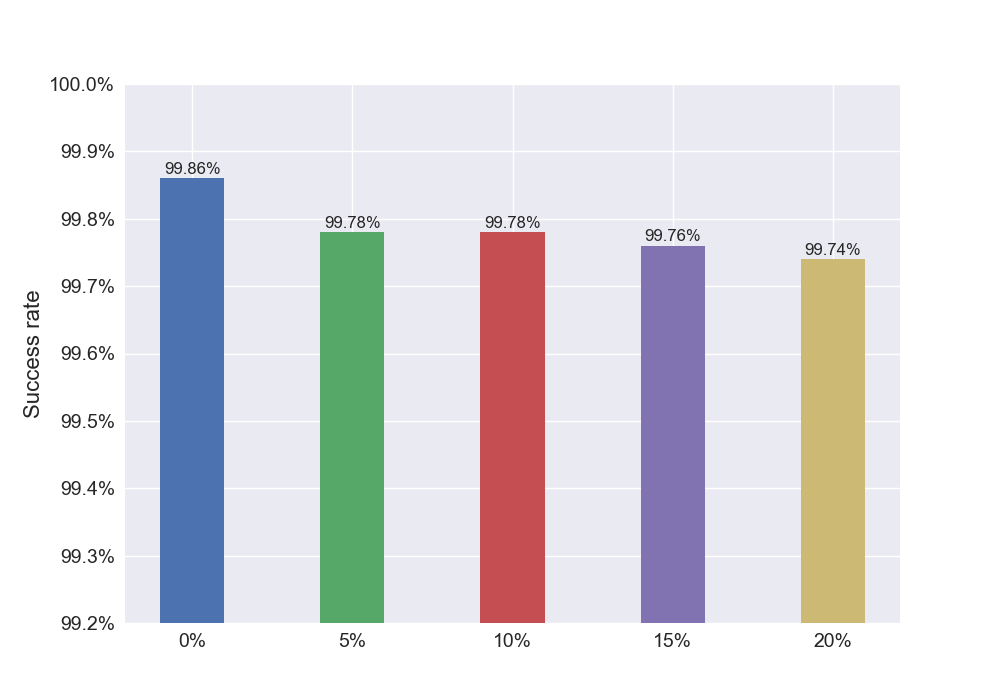}
	} 
	\subfigure[Organick \textit{et al.}\cite{organick2018random}]{
		\includegraphics[width=0.5\textwidth]{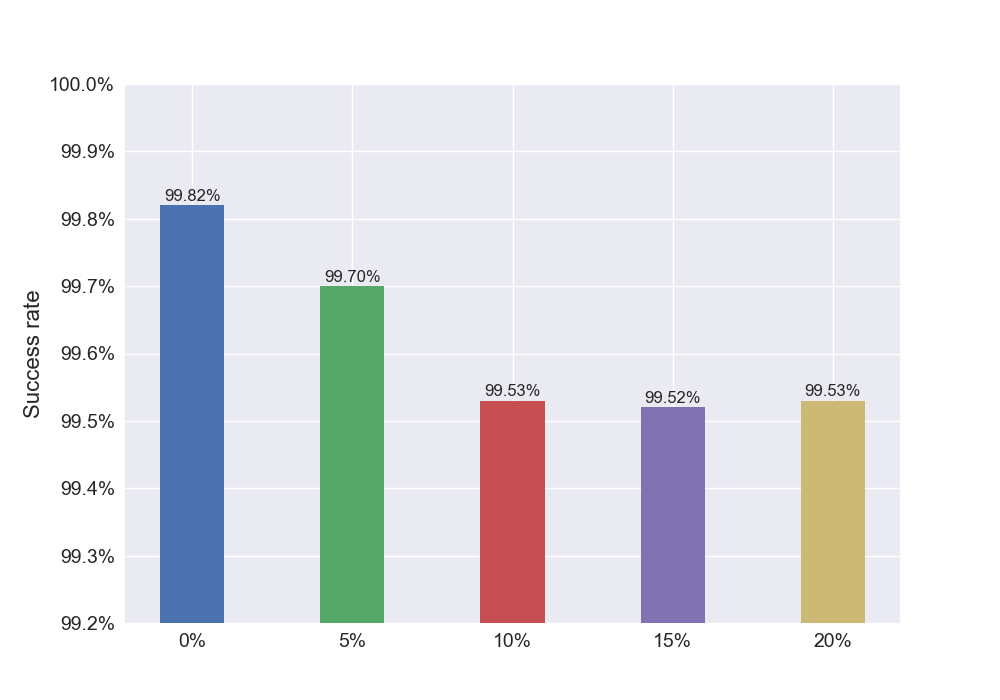}
	} 
	\subfigure[ Chandak \textit{et al.}\cite{chandak2019improved}]{
		\includegraphics[width=0.5\textwidth]{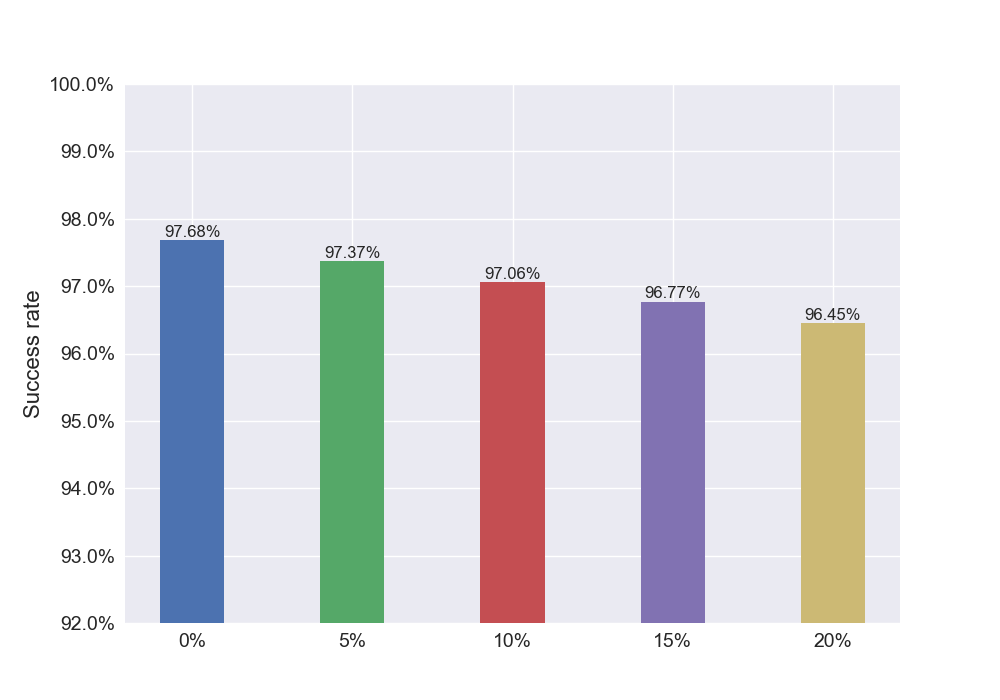}
	} 
	\captionsetup{labelfont=bf}
	\caption{Changes of reconstruction success with respect to the contamination level ranging from $0\%$ to $20\%$ using the proposed RRCC-DNN, on three datasets. }
	\label{fig:result1}
\end{figure}

\begin{figure}[htbp]
	\centering
	\subfigure[Erlich \textit{et al.}\cite{erlich2017dna}]{
		\includegraphics[width=0.48\textwidth]{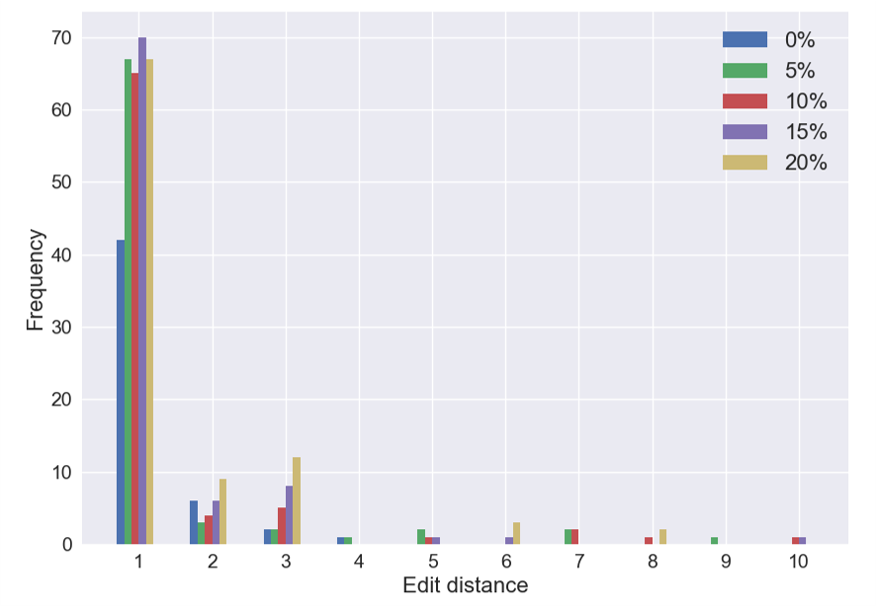}
	} 
	\subfigure[Organick \textit{et al.}\cite{organick2018random}]{
		\includegraphics[width=0.48\textwidth]{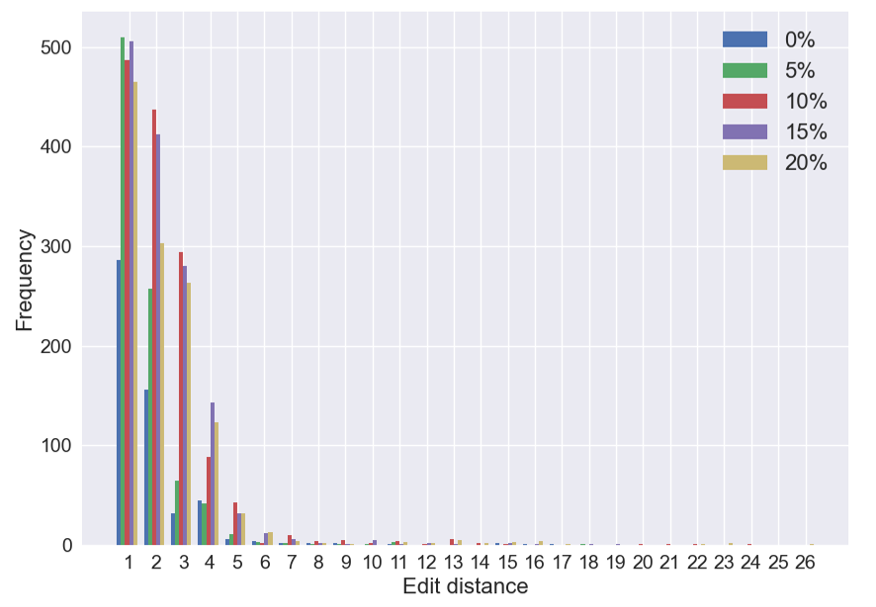}
	} 
	\subfigure[Chandak \textit{et al.}\cite{chandak2019improved}]{
		\includegraphics[width=0.48\textwidth]{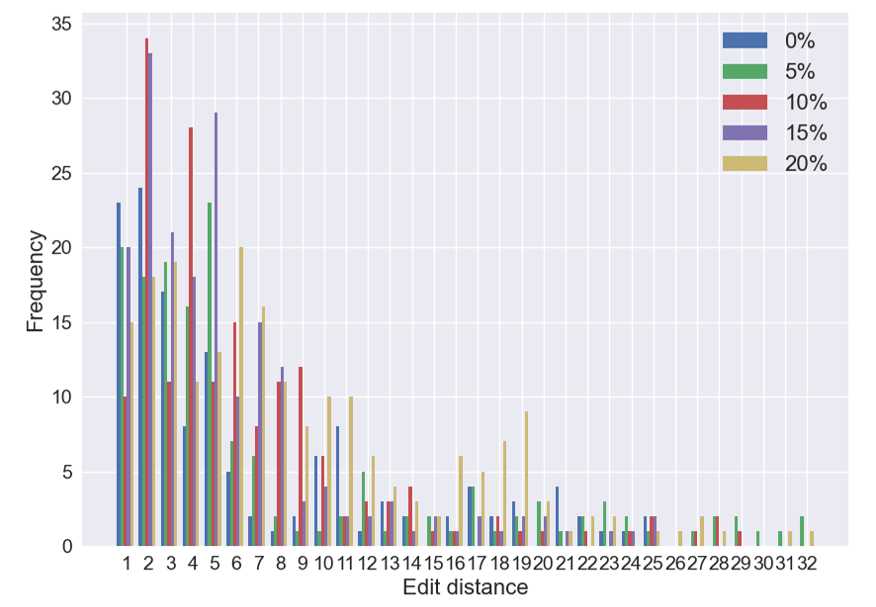}
	} 
	\captionsetup{labelfont=bf}
	\caption{Frequency histograms of the edit distance measured between the wrong prediction and the corresponding cluster reference.}
	\label{fig:error}
\end{figure}

\subsection{Evaluation metric and Comparative methods }
The effectiveness of the proposed method is evaluated by comparing with three state-of-the-art sequence reconstruction methods, where
the performance is evaluated by the success rate, given by
\begin{equation}
	success\ rate=\frac{\# \{predicted \ sequence = input \ reference\}}{\#  \{input \ reference\}}.
\end{equation}
In this formula, a sequence will contribute to the success rate only if it is perfectly reconstructed without error at every index position. 
\begin{itemize}
	\item {\textbf{Iterative Reconstruction} \cite{sabary2020reconstruction}: This algorithm uses multiple methods to revise strands from clusters and return the candidate sequence most likely to be the original reference. 
	The error vectors majority algorithm is used to correct insertion and substitution errors,  while the pattern-path algorithm is applied to correct deletion errors.}
	\item {\textbf{Divider BMA} \cite{sabary2020reconstruction}: This BMA-based algorithm divides the received clusters into three sub-clusters by their length. The majority voting is applied to the sequences of correct length. Then deletion and insertion error corrections are performed on the sub-clusters with shorter and larger sequence lengths, respectively.}
	\item {\textbf{BMA Lookahead} \cite{gopalan2018trace}: This is an improved algorithm of the BMA method. For sequences whose current symbol does not match the majority of symbols, a "prior window" looking at the next two (or more) symbols is used.}
\end{itemize}

\subsection{Results anaysis}

We report the reconstruction success rates of the proposed RRCC-DNN at different contamination levels $\{0\%, 5\%, 10\%, 15\%, 20\%\}$ on the testing set for all the three datasets, as shown in Figure~\ref{fig:result1}. 
For Erlich \textit{et al.}\cite{erlich2017dna}, the success rates reach 99.86\%, 99.78\%, 99.78\%, 99.76\%, and 99.74\%, corresponding to 51, 78, 79, 87, and 93 wrong predictions out of 36000 clusters. 
The success rates are 99.82\%, 99.70\%, 99.53\%, 99.52\%, and 99.58\% with 540, 897, 1391, 1408, and 1230 wrong predictions out of 296325 clusters for 
 {Organick \textit{et al.} \cite{organick2018random}.
 On the third dataset Chandak \textit{et al.}\cite{chandak2019improved}, the numbers are  97.68\%, 97.37\%, 97.06\%, 96.77\%, and 96.45\% with 136, 154, 172, 189, and 208 wrong predictions out of 5853 testing clusters. 
On all three datasets, the performance of the RRCC-DNN model remains stable with a slight decrease in success rate, as the proportion of contaminated strands gradually augmented even to 20\%.
This confirms the robustness and stability of the proposed RRCC-DNN model in presence of contaminated sequences.
Notice that the success rates are relatively low on the third dataset at all the contamination levels. It is due to the higher IDS error rates, as well as the mismatch in sequence lengths. 

\subsubsection{Wrong prediction analysis} Figure \ref{fig:error} illustrates the frequency histograms of the edit distance, which is measured between the incorrectly predicted sequence and the corresponding cluster reference. 
As observed, most of the wrong predictions have a small edit distance to their original reference, meaning that almost every sequence position can be correctly predicted by the proposed model, even if the cluster is not perfectly reconstructed.

\subsubsection{Impact of cluster size}

\begin{figure}[htbp]
	\centering
	\includegraphics[scale=0.46]{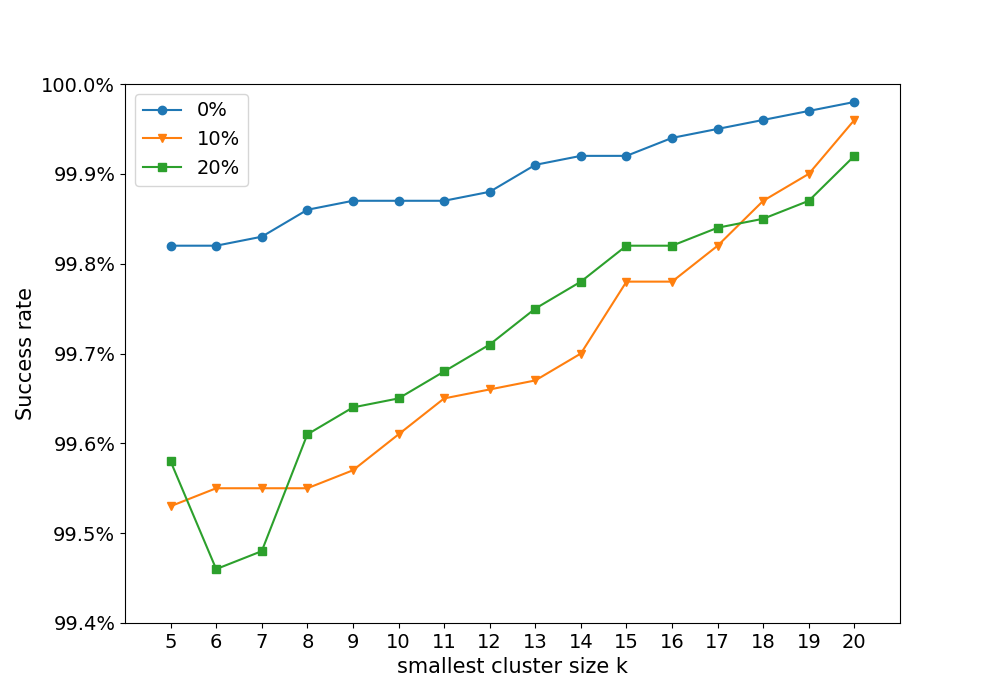}
	\captionsetup{labelfont=bf}
	\caption{Changes of the success rate in terms of the smallest cluster size $k$, namely the cluster size of the dataset ranges from $k$ to 30, on dataset~Organick \textit{et al.}\cite{organick2018random}.}
	\label{fig:cluster}
\end{figure}
We investigate how the smallest cluster size $k$ in a dataset affects the reconstruction success rate, under varying contamination conditions. 
The results are given in Figure \ref{fig:cluster}. 
As observed, for all contamination rates, the success rate increases with the increase of the smallest cluster size $k$ in a dataset. 
With $k=20$, the success rates achieves  99.98\%, 99.96\%, and 99.89\% under the contamination levels 0\%, 10\% and 20\%, respectively.

\subsection{Comparative study}
Figure \ref{fig:compare} reports the success rate obtained on all three datasets at varying contamination levels, by using the proposed RRCC-DNN, and other three sequence reconstruction strategies. 
We observe that on the dataset Chandak \textit{et al.} \cite{chandak2019improved},  the proposed RRCC-DNN always provides the best results at all contamination levels. This is because most reads in this dataset has a larger length than that of the original reference, signifying a high IDS error rate.
Thanks to the Conformer-Encoder module, our model can efficiently capture the general IDS error patterns, thus resilient to such position shifts within the strands. 
The BMA divider \cite{sabary2020reconstruction} fails on this dataset.

When the dataset is not contaminated at all, our method can provide comparable reconstruction results to state-of-the-art methods. 
On the first two datasets, the resulting success rates by RRCC-DNN are lower than that of Iterative Reconstruction~\cite{sabary2020reconstruction} and BMA Lookahead~\cite{gopalan2018trace}, but slightly higher than in BMA divider~\cite{sabary2020reconstruction}. 

The advantages of the proposed RRCC-DNN become increasingly evident as the proportion of contaminated sequences gradually augments in the dataset. 
For example, when the contamination proportion is increased to 10\% in Erlich \textit{et al.} \cite{erlich2017dna} data, the performance decreases in RRCC-DNN, Iterative Reconstruction\cite{sabary2020reconstruction}, BMA divider~\cite{sabary2020reconstruction}, and BMA Lookahead\cite{gopalan2018trace} are 0.1\%, 0.15\%, 0.37 and 0.22\%, respectively. Compared to its counterparts, the proposed method is least affected by cluster contamination. 
With 10\% contamination, the RRCC-DNN is second to Iterative Reconstruction~\cite{sabary2020reconstruction} by 0.06\% on dataset Erlich \textit{et al.} \cite{erlich2017dna}, and second to BMA Lookahead\cite{gopalan2018trace} by 0.07\% on the second data. 
When the contamination proportion reaches 15\%, the RRCC-DNN outperforms all the comparing methods on all three datasets. As the contamination proportion continues to increase, such advantages over other methods in terms of success rate become significant. 

The above discussions demonstrate the good behavior and robustness of the proposed RRCC-DNN in presence of contaminated sequences, as well as IDS errors within clusters. 
 
%

\begin{figure*}[htbp]
	\centering
	\subfigure[Contamination level: 0\%]{
		\includegraphics[width=0.48\textwidth]{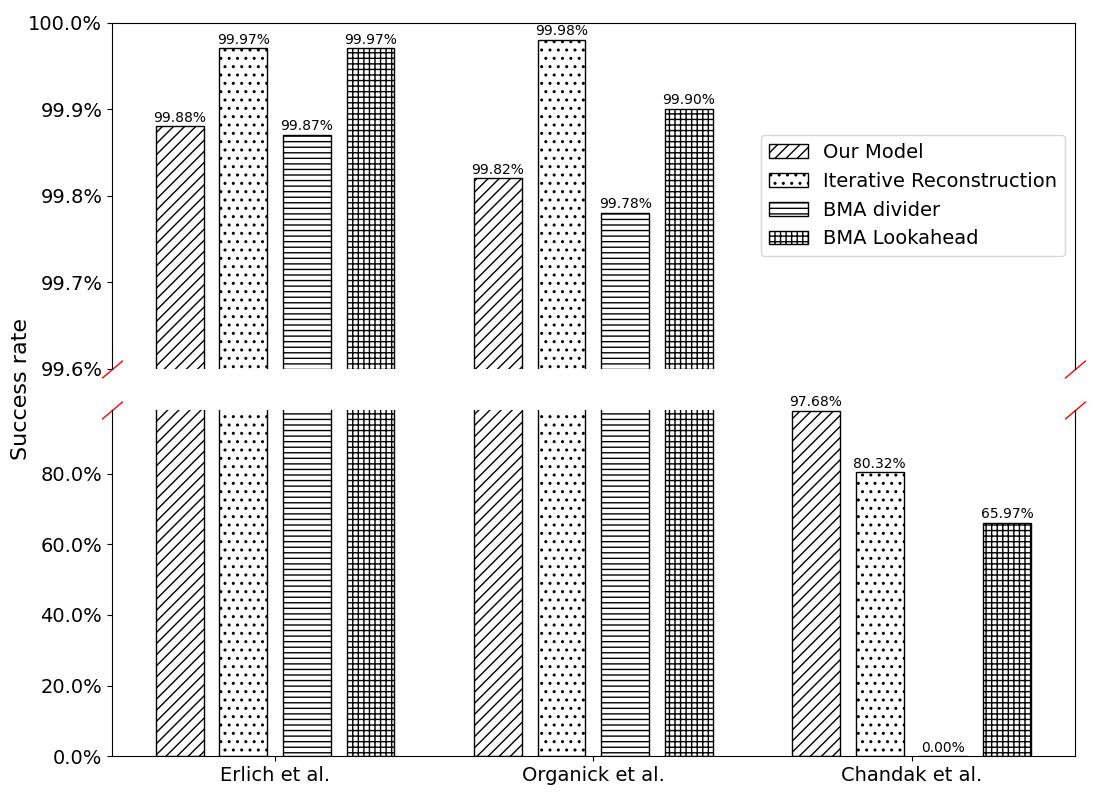}
	} 
	\subfigure[Contamination level: 10\%]{
		\includegraphics[width=0.48\textwidth]{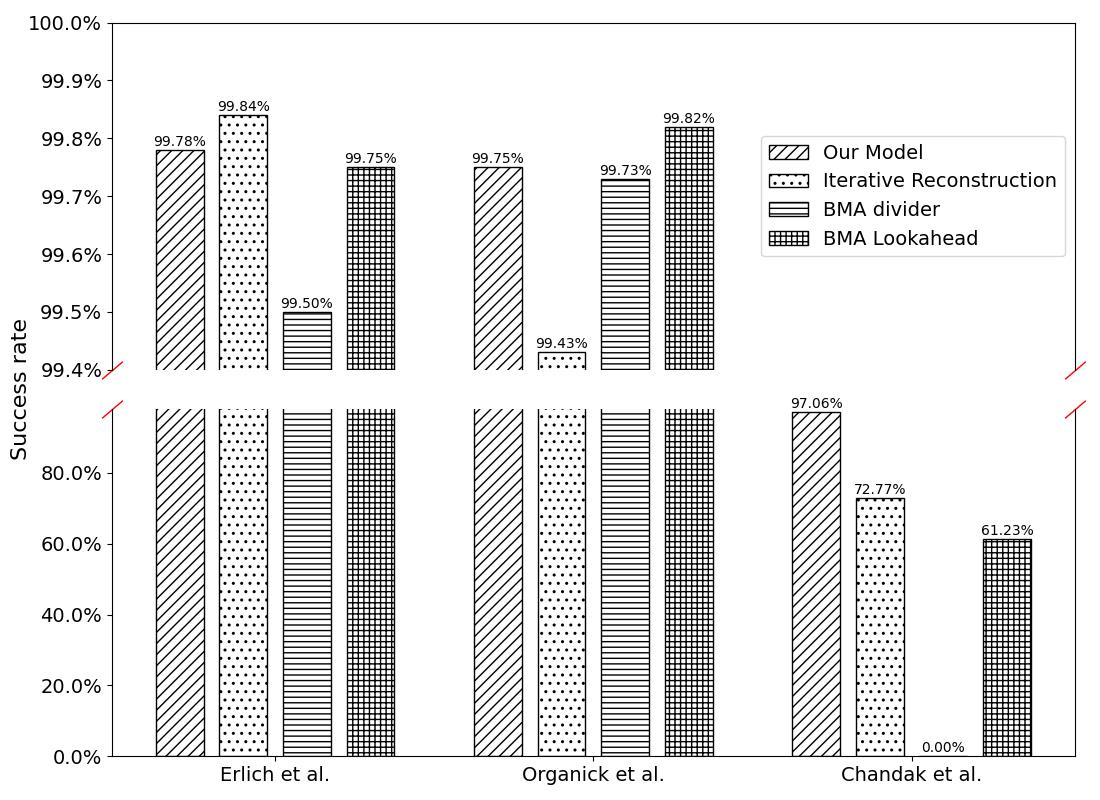}
	} 
	\subfigure[Contamination level: 15\%]{
		\includegraphics[width=0.48\textwidth]{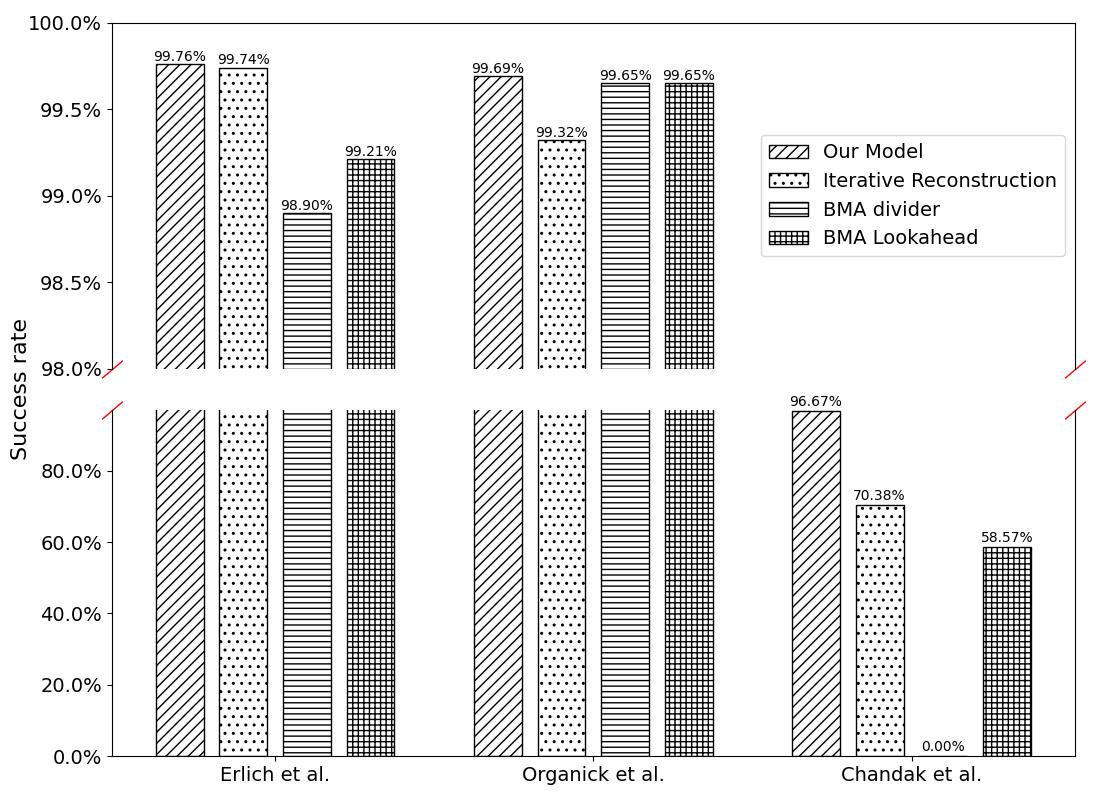}
	}
	\subfigure[Contamination level: 20\%]{
		\includegraphics[width=0.48\textwidth]{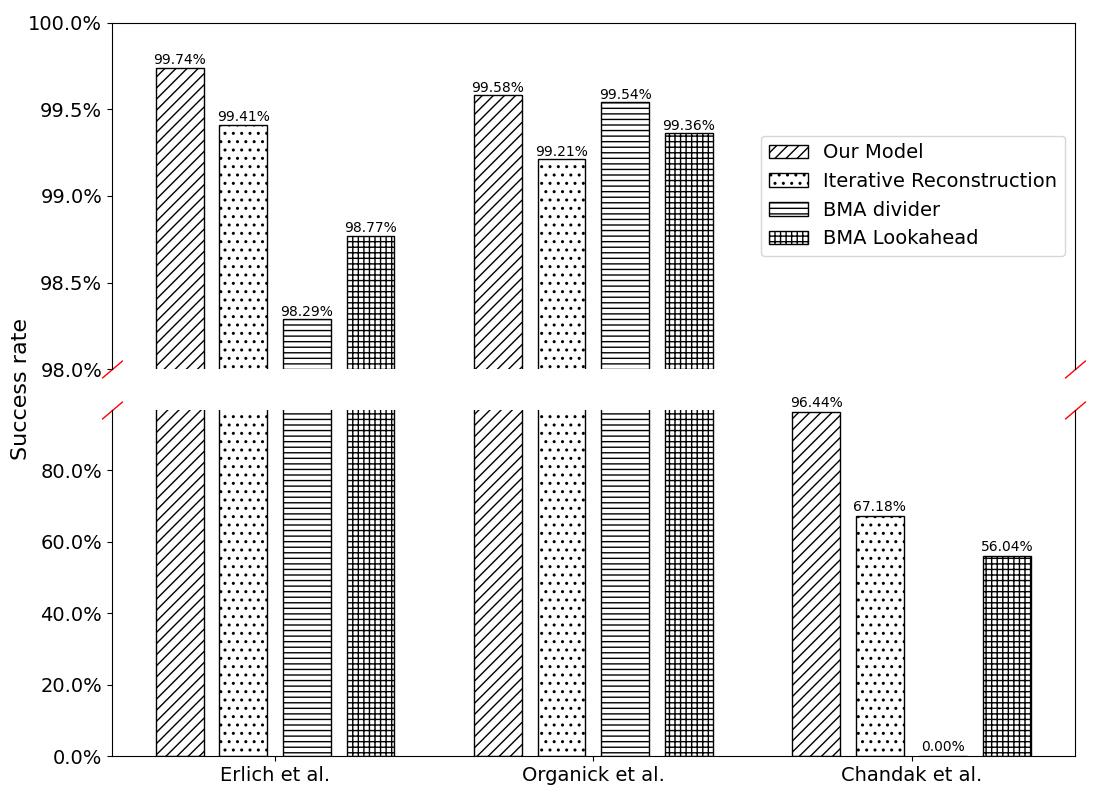}
	}
	\captionsetup{labelfont=bf}
	\caption{Comparison of success rate using the proposed RRCC-DNN, iterative reconstruction \cite{sabary2020reconstruction}, BMA divider~\cite{sabary2020reconstruction} and BMA lookahead \cite{gopalan2018trace}, on three datasets at contamination levels \%0, 10\%, 15\% and 20\%.}
	\label{fig:compare}
\end{figure*}

\subsection{Ablation Study}

\begin{table*}[htbp]
	\centering
	\captionsetup{labelfont=bf}
	\caption{{Ablation Study.} }
	\label{result4}
	\begin{tabular}{cccccccccc}
		\toprule
		\textbf{Model Architecture} & \multicolumn{9}{c}{\textbf{Success rate}} \\
		\midrule
		\multicolumn{1}{c}{} & \multicolumn{3}{c}{\textbf{Erlich \textit{et al.} \cite{erlich2017dna}}} & \multicolumn{3}{c}{\textbf{Organick \textit{et al.} \cite{organick2018random}}} & \multicolumn{3}{c}{\textbf{Chandak \textit{et al.} \cite{chandak2019improved}} } \\
		\cmidrule{2-10}    \multicolumn{1}{c}{} & \textbf{0\%} & \textbf{10\%} & \textbf{20\%} & \multicolumn{1}{c}{\textbf{0\%}} & \textbf{10\%} & \textbf{20\%} & \textbf{0\%} & \textbf{10\%} & \textbf{20\%} \\
		\cmidrule{2-10}    \textbf{Our Model} & \textbf{99.86\% } & \textbf{99.78\%} & \textbf{99.74\%} & \multicolumn{1}{c}{\textbf{99.82\%}} & \textbf{99.53\%} & \textbf{99.58\%} & \multicolumn{1}{c}{\textbf{97.68\%}} & \textbf{97.06\%} & \textbf{96.44\%} \\
		\textbf{-Attention} & 70.72\%  & 63.76\% & 43.45\% & \multicolumn{1}{c}{ 80.65\%} & 72.49\% & 57.53\% & \multicolumn{1}{c}{67.89\%} & 65.98\% & 62.79\% \\
		\textbf{-Attention+Normalization} & 99.21\% & 98.56\% & 96.34\% & \multicolumn{1}{c}{ 99.81\%} & 99.12\% & 97.51\% & \multicolumn{1}{c}{97.68\%} & 95.33\% & 90.12\% \\
		\textbf{-Conformer+Transformer} & 99.57\% & 99.32\% & 99.00\% & 99.15\% & 99.12\% & 98.47\% & 97.45\% & 96.78\% & 96.28\% \\
		\bottomrule
	\end{tabular}%
	
\end{table*}%
The ablation study is designed to demonstrate the necessity of the Attention Module and the effectiveness of the Conformer-Encoder. 
To this end, we first remove the attention mechanism in Attention Module and directly feed the model with the summation of all the input sequences within a cluster. As shown in Table \ref{result4}, the resulting model performs poorly on all datasets with varying contamination levels.  

We further impose an equal, normalized weight on every input strand. 
As seen from Figure \ref{result4}, the resulting reconstruction performance is always inferior to our proposed model, especially when severe contamination is present in the dataset. 
The gaps in success rate between the two models are up to  3.4\%, 2.07\%, and 6.32\% on three datasets when the contamination rate reaches 20\%.
In the case without additional contamination sequences, this model achieves similar results compared to our model. 

Finally, we modify our model by replacing the Conformer block with a Transformer block. The difference is that Conformer has a convolution module and a pair of Feed-Forward modules, while Transformer has only one Feed-Forward module~\cite{vaswani2017attention}. As shown in Table \ref{result4}, the performance of the latter model is satisfactory but inferior to our model, demonstrating the effectiveness of the proposed Conformer-Encoder. 

\section{Conclusion}~\label{sec5}
In this paper, we proposed a DNN-based multi-read reconstruction model for DNA storage, which is robust to noisy reads with IDS errors, and more importantly resilient to the contaminated sequences introduced during the DNA storage process. 
The proposed network has an encoder-decoder architecture with three pivotal components. 
The Attention Module suppresses the effect of contaminated sequences on the reconstruction, by automatically scoring the strands within the cluster and generating a representative, weight-averaged feature for subsequent tasks.  
The Conformer-Encoder has a sandwich structure and tackles most of the IDS errors within a cluster thanks to its advanced feature extraction capacity.
The single-layer LSTM-decoder finally predicts the reference DNA of the input cluster. 
We prove the effectiveness and robustness of the proposed RRCC-DNN on three next-generation sequencing datasets through a series of comparative experiments, where different levels of contamination caused by various factors during the process of DNA storage are simulated. 
The ablation study is also provided to verify the necessity of the attention mechanism and the conformer block in the proposed model. Future works will focus on adapting the proposed sequence reconstruction model to the Nanopore sequencing data with higher error rates. 

\bibliographystyle{IEEEtran}

\bibliography{ref}

\end{document}